\documentclass[final,3p,times,twocolumn]{elsarticle}

\usepackage{amsmath,amssymb}
\usepackage{lineno,hyperref,float}
\usepackage{textcomp}
\usepackage{color}
\usepackage{ulem}
\modulolinenumbers[5]

\journal{Journal of \LaTeX\ Templates}

\newcommand{\Deltay}{\Delta Y}

\newcommand{\Fe}{\rm{${}^{55}$Fe} }

\newcommand{\MnKa}{Mn--K$\alpha$ }

\begin{document}

\begin{frontmatter}

\title{Experimental studies on the charge transfer inefficiency of CCD developed
for the soft X-ray imaging telescope Xtend aboard the XRISM satellite}


%
%
\author[a]{Yoshiaki Kanemaru \corref{correspondingauthor}}
\author[a]{Jin Sato}
\author[a]{Toshiyuki Takaki}
\author[a]{Yuta Terada}
\author[a,b]{Koji Mori}
\author[c]{Mariko Saito}
\author[c,d]{Kumiko K. Nobukawa}
\author[e]{Takaaki Tanaka}
\author[e]{Hiroyuki Uchida}
\author[f,g]{Kiyoshi Hayashida}
\author[f,g]{Hironori Matsumoto}
\author[f,g]{Hirofumi Noda}
\author[f]{Maho Hanaoka}
\author[f]{Tomokage Yoneyama}
\author[f]{Koki Okazaki}
\author[f]{Kazunori Asakura}
\author[f]{Shotaro Sakuma}
\author[f]{Kengo Hattori}
\author[f]{Ayami Ishikura}
\author[e]{Yuki Amano}
\author[e]{Hiromichi Okon}
\author[e]{Takeshi G. Tsuru}
\author[h]{Hiroshi Tomida}
\author[i]{Hikari Kashimura}
\author[i]{Hiroshi Nakajima}
\author[j]{Takayoshi Kohmura}
\author[j]{Kouichi Hagino}
\author[k]{Hiroshi Murakami}
\author[l]{Shogo B. Kobayashi}
\author[a]{Yusuke Nishioka}
\author[a]{Makoto Yamauchi}
\author[a]{Isamu Hatsukade}
\author[m]{Takashi Sako}
\author[m]{Masayoshi Nobukawa}
\author[n]{Yukino Urabe}
\author[n]{Junko S. Hiraga}
\author[o]{Hideki Uchiyama}
\author[p]{Kazutaka Yamaoka}
\author[b]{Masanobu Ozaki}
\author[b,q]{Tadayasu Dotani}
\author[f]{Hiroshi Tsunemi}
\cortext[correspondingauthor]{Corresponding author}


\address[a]{Faculty of Engineering, University of Miyazaki, 1-1 Gakuen Kibanadai Nishi, Miyazaki 889-2192, Japan}
\address[b]{Japan Aerospace Exploration Agency, Institute of Space and Astronautical Science, 3-1-1 Yoshino-dai, Chuo-ku, Sagamihara, Kanagawa 252-5210, Japan}
\address[c]{Department of Physics, Faculty of Science, Nara Women's University, Kitauoyanishi-machi, Nara, Nara 630-8506, Japan}
\address[d]{Department of Physics, Kindai University, 3-4-1 Kowakae, Higashi-Osaka, Osaka 577-8502, Japan}
\address[e]{Department of Physics, Kyoto University, Kitashirakawa Oiwake-cho, Sakyo, Kyoto, Kyoto 606-8502, Japan}
\address[f]{Department of Earth and Space Science, Osaka University, 1-1 Machikaneyama-cho, Toyonaka, Osaka 560-0043, Japan}
\address[g]{Project Research Center for Fundamental Sciences, Osaka University, 1-1 Machikaneyama-cho, Toyonaka, Osaka 560-0043, Japan}
\address[h]{Japan Aerospace Exploration Agency, Institute of Space and Astronautical Science, 2-1-1, Sengen, Tsukuba, Ibaraki 305-8505, Japan}
\address[i]{College of Science and Engineering, Kanto Gakuin University, 1-50-1 Mutsuurahigashi, Kanazawa-ku, Yokohama, Kanagawa 236-8501, Japan}
\address[j]{Department of Physics, Tokyo University of Science, 2641 Yamazaki, Noda, Chiba 270-8510, Japan}
\address[k]{Faculty of Liberal Arts, Tohoku Gakuin University, 2-1-1 Tenjinzawa, Izumi-ku, Sendai, Miyagi 981-3193, Japan}
\address[l]{Department of Physics, Tokyo University of Science, 1-3, Kagurazaka, Sinjuku-ku, Tokyo 162-0825, Japan}
\address[m]{Faculty of Education, Nara University of Education, Takabatake-cho, Nara, Nara 630-8528, Japan}
\address[n]{Department of Physics, Kwansei Gakuin University, 2-2 Gakuen, Sanda, Hyogo 669-1337, Japan}
\address[o]{Science Education, Faculty of Education, Shizuoka University, 836 Ohya, Suruga-ku, Shizuoka 422-8529, Japan}
\address[p]{Department of Physics, Nagoya University, Furo-cho, Chikusa-ku, Nagoya, Aichi 464-8602, Japan}
\address[q]{Department of Space and Astronautical Science, School of Physical Sciences, SOKENDAI (The Graduate University for Advanced Studies), 3-1-1 Yoshino-dai, Chuou-Ku, Sagamihara, Kanagawa 252-5210, Japan}

\begin{abstract}
We present experimental studies on the charge transfer inefficiency (CTI) of charge-coupled device (CCD) developed for the soft X-ray imaging telescope, Xtend, aboard the XRISM satellite.  The CCD is equipped with a charge injection (CI) capability, in which sacrificial charge is periodically injected to fill the charge traps. By evaluating the re-emission of the trapped charge observed behind the CI rows, we find that there are at least three trap populations with different time constants. The traps with the shortest time constant, which is equivalent to a transfer time of approximately one pixel, are mainly responsible for the trailing charge of an X-ray event seen in the following pixel. A comparison of the trailing charge in two clocking modes reveals that the CTI depends not only on the transfer time but also on the area, namely the imaging or storage area. We construct a new CTI model by taking into account both transfer-time and area dependence. This model reproduces the data obtained in both clocking modes consistently. We also examine apparent flux dependence of the CTI observed without the CI technique. The higher incident X-ray flux is, the lower the CTI value becomes. It is due to a sacrificial charge effect by another X-ray photon. This effect is found to be negligible when the CI technique is used.

\end{abstract}


\end{frontmatter}

\section{Introduction}

The X-ray Imaging and Spectroscopy Mission (XRISM) is planned to be launched in the early 2020s, and will carry the soft X-ray imaging telescope, Xtend \cite{tashiro2018}. Xtend consists of the X-ray Mirror Assembly (XMA) and the Soft X-ray Imager (SXI) \cite{hayashida2018}. The SXI is an X-ray charge-coupled device (CCD) camera and employs four back-illuminated devices arranged in a $2 \times 2$ array, which covers a $38' \times 38'$ field-of-view with a total imaging area size of $62~\mathrm{mm} \times 62~\mathrm{mm}$.
The energy range is 0.4--13~keV and the energy resolution defined as full width at half maximum (FWHM) is required to be less than 200 and 250~eV for 6~keV X-rays 
just after and three years after the launch, respectively.

A CCD is a charge transfer device, and charge transfer is inevitably accompanied
by charge loss. This is because signal charge is trapped by lattice defects in the transfer channel, 
which could be generated by various mechanisms including CCD process, radiation damage, and so on.
Therefore, the charge transfer inefficiency (CTI), the
fraction of charge loss per transfer, is an important measure in evaluating 
CCDs. In addition to trap populations and densities, the CTI generally depends on several
operation parameters: transfer time \cite{gendreau1993}, operating
temperature \cite{sembay2004,grant2006,mori2013}, amount of transfer charge
\cite{townsley2002,lamarr2004,nobukawa2014}, and the existence of ``sacrificial charge''
\cite{gendreau1995,grant2003,bautz2004,nakajima2008,uchiyama2009,todoroki2012}. In the case of a photon-counting X-ray CCD, which serves as a spectrometer as well as an
imager, the charge loss during transfer results in misidentification of
an incident photon energy. Understanding and appropriate modeling of 
the charge loss during transfer in our operation conditions are 
essential to fully derive its spectroscopic performance.

In this paper, we present experimental studies on the CTI of X-ray CCDs developed for the XRISM SXI. 
We demonstrate the time constants of the charge traps, the area dependence of the CTI, an updated CTI model, and the flux dependence of the CTI in this order.

\section{Specifications, Operation, and Experiments}
\label{sec:Exp}

\begin{figure}
	\centering
	\includegraphics[width=1\linewidth]{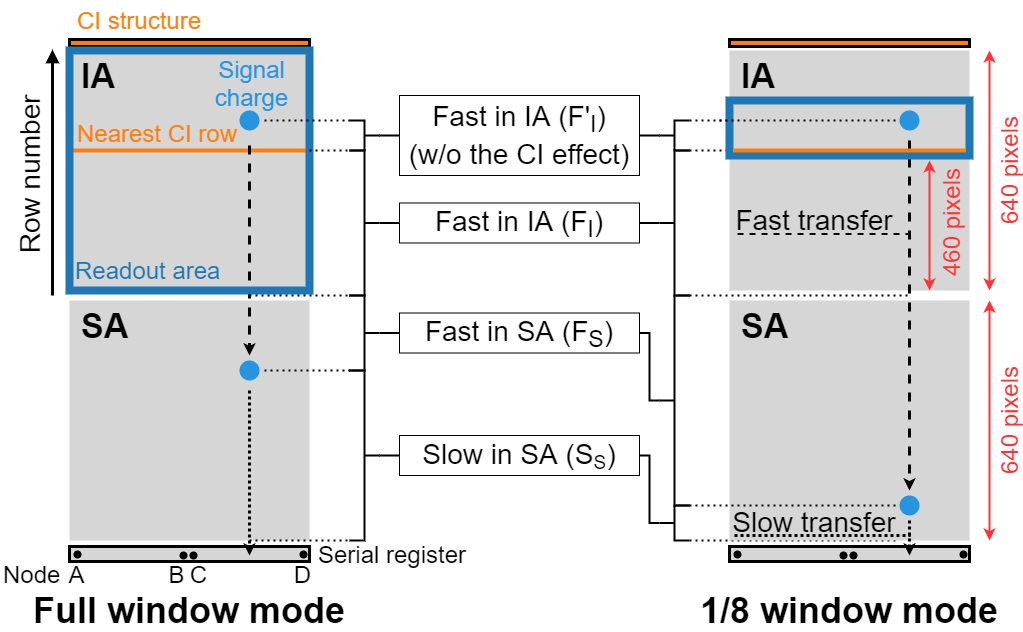}
	\caption{Schematic view of the charge transfer in the full and 1/8 window modes.}
	\label{fig:transferCond}
\end{figure}

We have developed flight model (FM) CCDs for the XRISM SXI with Hamamatsu Photonics K.K. The basic specifications were inherited from the CCDs developed for the SXI aboard Hitomi \cite{tanaka2018}; however, we paid particular attention to improvements in the optical-blocking performance \cite{uchida2020} and radiation tolerance \cite{kanemaru2019}. Here, we briefly summarize the CCD specifications that are relevant to this study. It is a buried p-channel CCD with a notch structure. The architecture is frame transfer. The pixel format is $1280 \times 1280$. The pixel sizes in the imaging area (IA) and the storage area (SA) are $24~\mu\mathrm{m} \times 24~\mu\mathrm{m}$ and 22--24$~\mu\mathrm{m}~\mathrm{(Horizontal;~H)} \times 16~\mu\mathrm{m}~\mathrm{(Vertical;~V)}$, respectively. One chip has four readout nodes, called A, B, C, and D (see \autoref{fig:transferCond}). The left and right halves of the IA are referred to as segments AB and CD, respectively, and are simultaneously read out with nodes A and C, respectively. Nodes B and D are used as a redundant option.
Our CCD is equipped with a charge injection (CI) structure on top of the columns, which injects sacrificial charge periodically to fill the traps. With this CI technique, an amount of charge corresponding to $\sim$~(3--8)$\times 10^{4}~\mathrm{e}^{-}$ is injected into every 160 rows. 
Applying $2\times2$ binning, we obtain a frame data with the effective pixel size and pixel format
of $48~\mu\mathrm{m} \times 48~\mu\mathrm{m}$ and $320~\mathrm{(H)} \times
640~\mathrm{(V)}$ for each segment, respectively. In what follows, our descriptions
are given in the unit of the binned pixel. The CI rows are then seen every 80 rows in the frame data.

We prepared two clocking modes for in-orbit operation: the full window mode and the 1/8 window mode. \autoref{fig:transferCond} shows a schematic view of the charge transfer in the two clocking modes. The full window mode is a nominal mode to read out the entire IA in 4~sec. The 1/8 window mode reads out one-eighth of the IA in 0.5~sec. The readout area position, 460 pixels above the bottom of the IA, was selected to cover the focal point of the XMA. This mode is designed for observations of bright point sources to reduce event pile-up and improve the time resolution at the expense of the readout area size. In both clocking modes, the transfer times are 57.6~$\mu$sec from the IA to the SA (hereafter referred to as fast transfer) and 5240~$\mu$sec per pixel from the SA to the serial register (hereafter referred to as slow transfer), respectively. Considering the transfer time, area, and CI effect, there are four kinds of transfers as follows:
\begin{enumerate}
	\item Fast transfer in the IA before reaching the nearest CI row ($\mathrm{F'}_{\mathrm{I}}$),
	\item Fast transfer in the IA after passing the nearest CI row ($\mathrm{F}_{\mathrm{I}}$),
	\item Fast transfer in the SA ($\mathrm{F}_{\mathrm{S}}$), and
	\item Slow transfer in the SA ($\mathrm{S}_{\mathrm{S}}$).
\end{enumerate}
The number of transfers to the serial register depends on the location of the X-ray
event and common in the two clocking modes. However, in the two clocking modes, the
ratios among the numbers of the four types of transfers differ, 
and therefore the amounts of charge loss can also differ.

We performed on-ground calibration experiments for all four FM CCDs, FM02-02, 09, 10, and 13, with the two clocking modes at \textminus110~\textcelsius{}, which is the initial operation temperature in orbit. 
We also performed supplemental experiments using the same type of a CCD at \textminus120~\textcelsius{}, to which the operation temperature could be lowered. 
The CCDs were irradiated with X-rays from \Fe in the experiments. From the frame data obtained, we extracted X-ray event data consisting of a $3 \times 3$ pixel island in the same manner as applied in the Hitomi SXI \cite{nobukawa2014}. The pulse-height of the pixels in each event island was calculated by subtracting the dark level. We used only grade-0 events, whose signal charge was observed in a single pixel alone.

\section{Analysis and Results}

In the following analysis, we used raw frame data obtained by the FM CCDs at
\textminus110~\textcelsius{} for the time constant measurement of the charge traps
(\autoref{sec:TC}). We analyzed the \Fe event data obtained by the FM CCDs at
\textminus110~\textcelsius{} for studies on the trailing charge
(\autoref{sec:evtTrail}) and CTI modeling (\autoref{sec:CTI}). The flux
dependence of the CTI (\autoref{sec:flux}) was examined by studying the \Fe events obtained with the same type of
CCD at \textminus120~\textcelsius{}. 
All the best-fit models in the following sections were determined by chi-square minimization with the Levenberg-Marquardt method.
\subsection{Measurement of time constants of the charge traps}
\label{sec:TC}
\begin{figure}[h]
	\centering \includegraphics[width=1\linewidth,trim={0 0 0 0},clip]{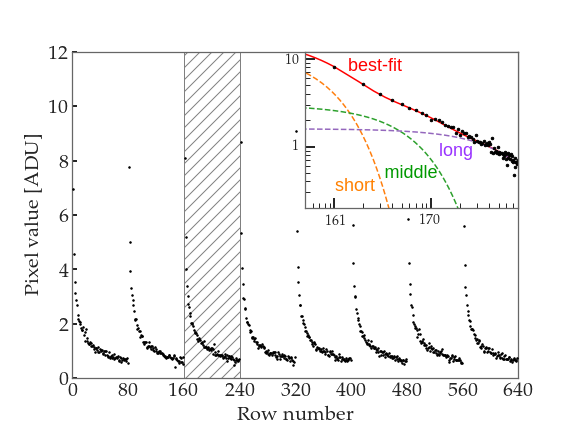}
	\caption{
		Pixel value of the frame data taken with the CI as a function of row number (black dots). The hatched area is shown in the upper-right inset on a logarithmic scale. The red solid line denotes the best-fit model while the orange, green, and purple dashed lines represent short, middle, and long components, respectively.}
	\label{fig:ciTrail}
\end{figure}
\autoref{fig:ciTrail} shows the pixel value of the frame data taken with the CI as a function of row number. 
The pixel value was calculated from a raw digital output with no X-rays irradiated in a pixel subtracted by the contribution of the bias voltage estimated from outputs in the horizontal overclock area.
We also subtracted the contribution of dark current estimated from the frame data taken without the CI. The data points shown in \autoref{fig:ciTrail} are averaged values in each row.
The figure indicates that the pixel value exponentially decays as a function of the distance from the preceding CI row. The amplitude of the decay component corresponds to the amount of re-emitted charge, $Q_{\mathrm{emit}}$, from the traps that were filled by a large amount of charge in the CI row \cite{prigozhin2000}. The pixel value can be well approximated by the following function:

\begin{align}
Q_{\mathrm{emit}}(\Deltay) ={}& 
\sum_{i}
a_{i}\exp\left(- \frac{\Deltay}{\tau_i} \right)~, \label{eq:threeExpModel}
\end{align}
where $\Deltay$ is the distance from the preceding CI row, and $\tau_i$ and $a_i$
are the time constant and normalization of each
component, respectively. 
We note that the ``time constant'' was defined in the unit of pixels for the convenience in the following analysis.
We found that there are at least three trap
populations having short, middle, and long time constants. 
They consist of approximately 1, 10, and 100 pixels (see the inset in
\autoref{fig:ciTrail}), which correspond to 57, 570, and 5700~$\mu$sec, respectively. 
We also noticed that they are almost independent of the segments or CCDs.

\begin{figure}[h]
	\centering
	\includegraphics[width=1\linewidth]{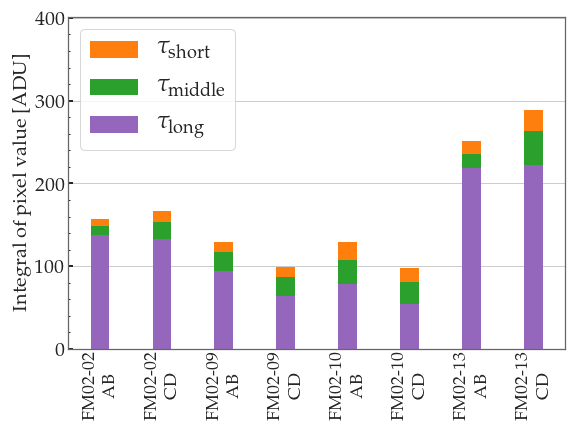}
	\caption{Integrals of the three components of
		\autoref{eq:threeExpModel} for each segment. The orange, green and purple
		bars denote those with short, middle, and long time constants,
		respectively.}  \label{fig:chargeAmount}
\end{figure}

\autoref{fig:chargeAmount} summarizes the integrals of the three components 
up to infinity for each segment. The differences between the segments may result from variations in the trap density in the channel and/or the amount of injected charge. This figure indicates that trap populations with the short, middle, and long time constants account for 4--17, 7--27, and 56\%--89\% of 
the total charge loss
proportional to the sum of integrals, respectively. 
As shown in \autoref{fig:ciTrail}, the re-emission from the trap with the short time constant
accounts for a large part of charge in the following pixel to a CI row.
Limiting the integration range up to the following pixel, the numbers
become 62--75, 17--22, and 5\%--19\%, respectively.

\subsection{Area dependence of the charge trail}
\label{sec:evtTrail}

\begin{figure}[h]
	\centering \includegraphics[width=1\linewidth,trim={0 0 0
		0pt},clip]{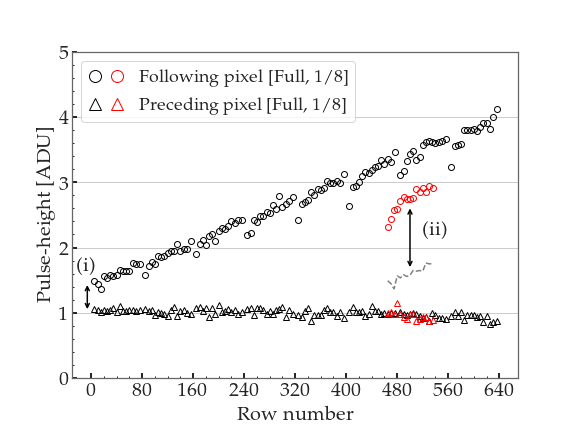}
	\caption{
		Averaged pulse-height of preceding (triangles) and following (circles) pixels of grade-0 events as a function of row number. The black and red symbols represent the data taken in the full and 1/8 window modes, respectively. The dashed line indicates the pulse-height of the full window mode data below the row number of 80, which is drawn to make it easier to compare the pulse-height with that of the 1/8 window mode data. See texts for the meaning of labels (i) and (ii).
	}
	\label{fig:evtTrail}
\end{figure}

In X-ray event data, where the amount of charge generated by an X-ray photon is significantly smaller than that injected into a pixel by the CI structure, re-emitted charge in one or two following pixels is only virtually observable in each event, which is the so-called charge trail. \autoref{fig:evtTrail} shows the averaged pulse-height of the preceding and following pixels of grade-0 events as a function of row number.
The pulse-height is calculated from a raw digital output subtracted by that with no X-rays irradiated, namely a dark level, so that it represents the amount of charge originated from X-rays. The practical difference between the pixel value used in \autoref{sec:TC} and the
pulse-height defined here is that the former contains re-emitted charge from CI rows
whereas the latter does not.
The pulse-height of the following pixel increases with row number,
namely the number of transfers, as the number of re-emissions into the following pixel also increases. 
The pulse-height of the preceding pixel is almost constant, which represents the amount of split charge to neighboring pixels below the threshold used in the event grading. The difference in pulse-height between the following and preceding pixels corresponds to the amount of ``trailing charge''. It almost linearly increases with row number, but discontinuously drops just after the CI rows. 
This is because the number of filled traps discontinuously increases just after the
CI row and gradually decreases with the time constants we measured; the amount of re-emitted charge, which is inversely proportional to the number of 
the traps filled by the injected charge, is accordingly lesser in the events generated closer to the CI row.

The amount of trailing charge at the bottom of the IA, labeled as (i) in \autoref{fig:evtTrail}, can be a measure of the CTI of $\mathrm{F_{S}}$ because of the dominance of $\mathrm{F_{S}}$ in the entire transfer. Similarly, the difference in the amount of trailing charge at the same $\Delta Y$ between the data of the 1/8 window mode and the first 80 pixel data of the full window mode, labeled as (ii) in \autoref{fig:evtTrail}, can be a measure of the CTI of $\mathrm{F_{I}}$ because the difference is solely attributed to the difference in the number of $\mathrm{F_{I}}$. In other words, the numbers of $\mathrm{F_{I}'}$, $\mathrm{F_{S}}$, and $\mathrm{S_{S}}$ are the same (\autoref{fig:transferCond}).

\begin{figure}
\centering
\includegraphics[width=1\linewidth]{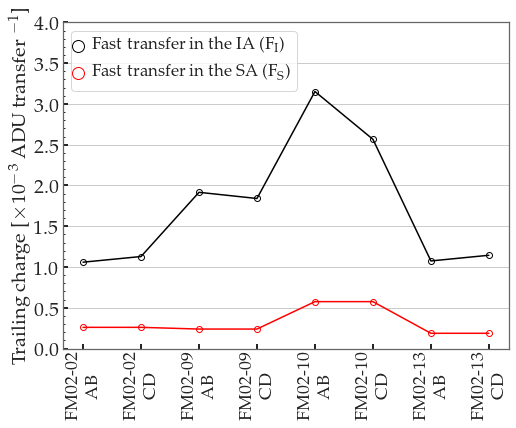}
\caption{Amount of trailing charge per transfer of $\mathrm{F_{S}}$ (red circle) and $\mathrm{F_{I}}$ (black circle) for all segments. The solid lines are guides for the eye.}
\label{fig:evtTrailSummary}
\end{figure}

\autoref{fig:evtTrailSummary} summarizes the amounts of the trailing charge per transfer of $\mathrm{F_{S}}$ and $\mathrm{F_{I}}$ for all segments. The former and the latter were derived from values (i) and (ii) by dividing by 640 and 460, respectively. The value (i) was estimated based on the difference in the intercepts of the two lines that were obtained by fitting each data set of the preceding and following pixels in the full window mode. The value (ii) was estimated at $\Delta Y > 20$, for which the value is almost constant. Although the transfer times of $\mathrm{F_{S}}$ and $\mathrm{F_{I}}$ are identical, the amount of trailing charge in the IA is several times larger than in the SA in all segments. This result clearly shows that the CTI depends on the transfer area. 

\subsection{CTI model considering area dependence}
\label{sec:CTI}

We cannot extract information about the trap with the
long-time constant from the following pixels, since the
amount of charge in the following pixels re-emitted by those
traps is much smaller than those from the traps with other
time constants.
To restore the initial pulse-height, we need to estimate the amount of the charge loss by a CTI model, which should well predict the pulse-height observed as a function of row number. In the case of the Hitomi SXI, we introduced a CTI model that distinguishes the CTIs between the fast and slow transfers but not between the imaging and storage areas \cite{nobukawa2014}. 
The CTI model describes the data obtained in the full window mode, whereas it is not yet verified if the same model with the same parameters can reproduce the data obtained in the 1/8 window mode.

As described in \autoref{sec:evtTrail}, the CTIs due to $\mathrm{F_{I}}$ and $\mathrm{F_{S}}$ are different even though their transfer times are equal. Following the CTI model shown in Ref.\cite{nobukawa2014}, the pulse-height observed, $\mathit{PH}$, is expressed as follows:

\begin{align}
\mathit{PH} =~ \mathit{PH}_\mathrm{0} 
&\cdot
(1 - c_{{\mathrm{F_{I}'}}})^{\Deltay}
\cdot
(1 - c_{{\mathrm{F_{I}}}})^{Y_{{\mathrm{F_{I}}}}} \notag \\
&\cdot
(1 - c_{{\mathrm{F_{S}}}})^{Y_{{\mathrm{F_{S}}}}} 
\cdot
(1 - c_{{\mathrm{S_{S}}}})^{Y_{{\mathrm{S_{S}}}}}\,, 
\label{eq:ctiModelSpeedArea}
\end{align}
where $\mathit{PH}_{\mathrm{0}}$ is the initial pulse-height, and $c_i$ and $Y_i$ are the CTI and the number of transfers, respectively. The index $i$ denotes the transfer time and area as shown in \autoref{fig:transferCond}. Among the four $c_i$ functions, $c_{{\mathrm{F_{I}'}}}$ has only a constant factor and the other three are functions of $\Deltay$ as follows:

\begin{equation}
c_i = c_{\mathrm{0}i}
\cdot 
\left\{ 1 - p_i
\cdot \exp 
\left(
- \frac{ \Deltay}{\tau_i}
\right)
\right\} ~~(i = {\mathrm{F_{I}}}, {\mathrm{F_{S}}}, {\mathrm{S_{I}}})\,, 
\label{eq:cti}
\end{equation}
where $c_{0i}$, $p_i$, and $\tau_i$ are the CTI values without the CI effect ($c_{0\,\mathrm{F_{I}}} = c_{{\mathrm{F_{I}'}}}$), probability that the injected charge fills a trap, and net time constant of re-emission in the unit of a transfer number, respectively.

\begin{figure}[h]
	\centering
	\includegraphics[width=1\linewidth]{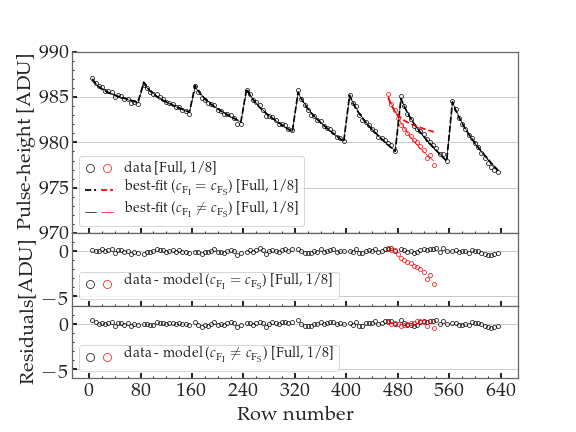}
	\caption{Pulse-height of the \MnKa line as a function of row number. The black and red symbols represent the data obtained in the full and 1/8 window modes, respectively. The dashed and solid lines show the best fits of the two CTI models with $c_{{\mathrm{F_{I}}}} = c_{{\mathrm{F_{S}}}} $ and $c_{{\mathrm{F_{I}}}} \ne c_{{\mathrm{F_{S}}}} $, respectively. The lower panels show the residuals from the models.
	}
	\label{fig:ctiModels}
\end{figure}

\autoref{fig:ctiModels} shows the pulse-height of the \MnKa line as a function of row number. 
The pulse-height periodically changes and follows a sawtooth shape because  the CTI depends on the fraction of traps filled by the injected charge. Therefore, the CTI is expressed as a function of the distance from the CI rows, as in \autoref{eq:cti}. \autoref{fig:ctiModels} also shows the residuals from the best fits of two CTI models: one is the previous model, wherein the CTI does not have an area dependence, $c_{{\mathrm{F_{I}}}} = c_{{\mathrm{F_{S}}}}$, and the other is the one described in \autoref{eq:ctiModelSpeedArea}. 
Both models reproduce the pulse-height observed in the full window mode well, but not that in the 1/8 window mode; 
the previous model underpredicts the charge loss. This would be explained as follows. In both models, $c_{{\mathrm{F_{S}}}}$
is determined almost solely by the charge loss in the first 80 pixels in the
full window mode because $\mathrm{F_{S}}$ dominates here. Following the
discussion in \autoref{sec:evtTrail}, 
we can determine the additional decrease in the pulse-height in the 1/8 window mode from the pulse-height of the first 80 pixel data in the full window mode 
because $\mathrm{F_{I}}$ dominates the difference of their transfers. As shown in \autoref{sec:evtTrail}, it is suggested to be 
$c_{{\mathrm{F_{I}}}} > c_{{\mathrm{F_{S}}}}$ in reality. Therefore, the
previous model, assuming $c_{{\mathrm{F_{I}}}} = c_{{\mathrm{F_{S}}}}$, predicts
smaller charge loss values than the actual values. In contrast, our model which
incorporates the area dependence can reproduce the data obtained in two different
clocking modes with identical parameters. For the XRISM SXI, we employ 
the CTI model that takes into account with both transfer-time and area dependence.
We confirmed that the energy resolutions after all of the corrections including the CTI model were smaller than 200~eV for 6~keV X-rays for all segments in the laboratory measurement; that is to say, FM CCDs met the requirement for the initial energy resolution.

\subsection{Flux dependence of CTI}
\label{sec:flux}
\begin{figure}[h]
		\centering
		\includegraphics[width=1\linewidth]{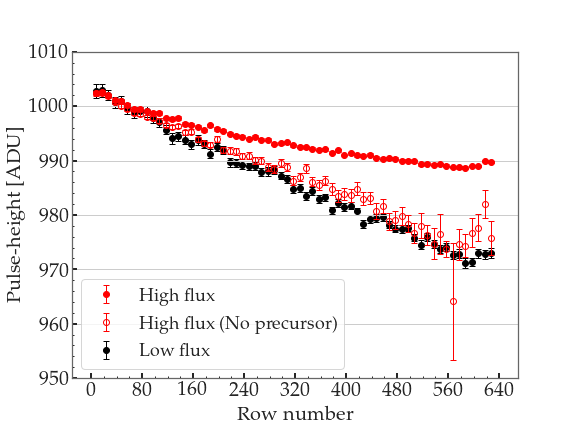}
		\caption{Pulse-height of the \MnKa line as a function of row number. The red and black filled circles represent the data taken with high and low X-ray fluxes, respectively. The red open circles denote the case using ``no precursor
			charge packet'' alone (see text for details).
	}
		\label{fig:ctiFluxHighLow}
\end{figure}
\autoref{fig:ctiFluxHighLow} shows the pulse-height of the \MnKa line as a function of row number. We obtained the data without the CI at two different X-ray fluxes, $\sim$1200 and $\sim$50 $\mathrm{counts~frame^{-1}~segment^{-1}}$. 
The former value is comparable with the count rate expected in the observation of the core region of the Perseus cluster, which is one of the brightest diffuse	sources in the X-ray sky.
Both data match each other well for very low row numbers. In the low-flux data, the pulse-height almost linearly decreases as a function of row number. In contrast, in the high-flux data, not only the pulse-height but also its local slope decrease as a function of row number. As a shallower slope means a smaller CTI value, it is indicated that the CTI value in the high-flux data apparently becomes smaller for higher row numbers. This phenomenon is expected, if a precursor charge packet, generated by another X-ray photon incident on a lower row number pixel in the the same column in the same	exposure of the X-ray photon of interest, serves as sacrificial charge. 
This chance coincidence becomes higher for higher flux data and a higher row number event. Using the high-flux data, we made the same plot but using the events with no precursor
charge packets alone, which is also shown in \autoref{fig:ctiFluxHighLow}. This ``no precursor charge packet'' plot matches well the low flux data plot, indicating that the above hypothesis is correct. This phenomenon obviously brings complexity to the interpretation of the results in spectroscopy because the calibration parameters determined at a certain flux value are not valid for different flux data sets.
In other words, the energy resolution is degraded and the peak energy is shifted by the correction for a certain flux that does not match observed data if the flux dependence is not negligible.

\begin{figure}[h]
		\centering
		\includegraphics[width=1\linewidth]{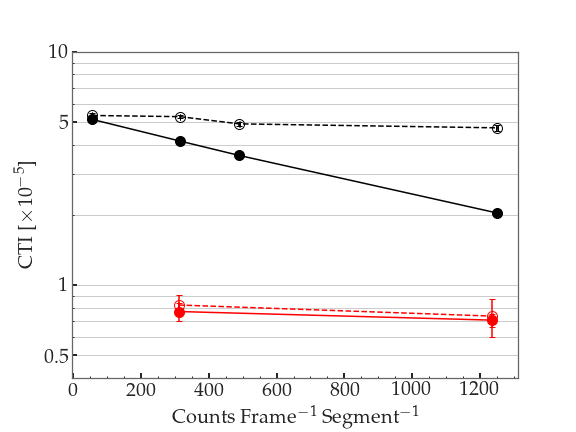}
		\caption{CTI as a function of incident X-ray flux. The red and black symbols represent the data taken with and without the CI, respectively. The open circles denote the case of using no precursor charge packets alone and filled circles represent the case of using all events. The dashed and solid lines are guides for the eye.}
		\label{fig:ctiRate}
\end{figure}

\autoref{fig:ctiRate} shows how the apparent CTI value depends on the incident X-ray flux. We obtained the data with various flux levels with and without the CI. We evaluated the CTI value by fitting the pulse-height of the \MnKa line vs. row number plots to
\begin{equation}
PH = PH_{\mathrm{0}}\,(1-\mathit{CTI})^Y,
\end{equation}
where $Y$, $\mathit{CTI}$, $PH$, and $PH_{\mathrm{0}}$ are the row number, CTI value, pulse-height of the \MnKa line at a given $Y$, and pulse-height of the \MnKa line at $Y=0$, respectively. As expected, in the data set without the CI, the CTI value becomes smaller for a higher flux, whereas it becomes almost constant, independent of the flux, in the case of using no precursor charge packets alone. 
In contrast, in the data set with the CI, the CTI value is almost constant regardless of the flux.
The same result was obtained using all events. 
The difference between the data sets with and without the CI can be explained by the fact that traps with a long time constant of $\sim$100 pixels are already filled by the injected charge.
 Because the XRISM SXI in orbit will operate only with the CI to fill traps by a large amount of charge, there is no need to consider the effects of precursor charge packets acting as sacrificial charge.

\section*{Summary}
We reported experimental studies on the CTI of FM CCD developed for 
Xtend aboard XRISM. The time constants of the charge traps were measured.
We found that there were at least three trap populations, and their time
constants in the unit of a pixel were approximately 1, 10, and 100 pixels. A
comparison of the amounts of trailing charge in the full and 1/8 window mode data
indicated that the charge loss due to a fast transfer in the IA is larger than
that in the SA. Considering the area dependence as well as the transfer-time dependence,
our CTI model was able to reproduce the charge loss of both full and 1/8 window mode data
with identical parameters. 
We confirmed that the spectroscopic requirement is met applying all of the corrections including the CTI model.
We also confirmed that the CTI depends on
the incident X-ray flux without the CI. In the case with the CI, the flux
dependence was found to be negligible. 
The new CTI model (\autoref{sec:CTI}) is directly useful for the XRISM SXI because it will be used to correct the CTI of the data obtained after the launch. The rest of our works are needed to build the model (\autoref{sec:TC}, \autoref{sec:evtTrail}) and verify the model to the incident flux (\autoref{sec:flux}).

\section*{Acknowledgements}
We thank all the XRISM Xtend team members.
We also thank the anonymous referee for useful comments and suggestions to improve the manuscript.
This work was supported by JSPS KAKENHI Grant Number
16H03983 (K.M.),
16J00548, 20K14491(K.K.N.),
20H01947 (H.N.),
18H01256 (H.N.),
and
17K14289 (M.N.)%
.

\bibliography{bib/hstd12_kanemaru_references.bib}
\end{document}